\documentclass[12pt,epsf,a4]{article}
\usepackage{graphicx}
\usepackage{amsmath}
\usepackage{amssymb}

\numberwithin{equation}{section}

\begin{document}

\def\eqrefc{\eqref}
\def\wt{\widetilde}
\def\a{\alpha}
\def\b{\beta}
\def\c{\varepsilon}
\def\d{\delta}
\def\e{\epsilon}
\def\f{\phi}
\def\g{\gamma}
\def\h{\theta}
\def\k{\kappa}
\def\l{\lambda}
\def\m{\mu}
\def\n{\nu}
\def\p{\psi}
\def\q{\partial}
\def\r{\rho}
\def\s{\sigma}
\def\t{\tau}
\def\u{\upsilon}
\def\v{\varphi}
\def\w{\omega}
\def\x{\xi}
\def\y{\eta}
\def\z{\zeta}
\def\D{\Delta}
\def\G{\Gamma}
\def\H{\Theta}
\def\L{\Lambda}
\def\F{\Phi}
\def\P{\Psi}
\def\S{\Sigma}

\def\o{\over}
\def\beq{\begin{eqnarray}}
\def\eeq{\end{eqnarray}}
\newcommand{\gsim}{ \mathop{}_{\textstyle \sim}^{\textstyle >} }
\newcommand{\lsim}{ \mathop{}_{\textstyle \sim}^{\textstyle <} }
\newcommand{\vev}[1]{ \left\langle {#1} \right\rangle }
\newcommand{\bra}[1]{ \langle {#1} | }
\newcommand{\ket}[1]{ | {#1} \rangle }
\newcommand{\EV}{ {\rm eV} }
\newcommand{\KEV}{ {\rm keV} }
\newcommand{\MEV}{ {\rm MeV} }
\newcommand{\GEV}{ {\rm GeV} }
\newcommand{\TEV}{ {\rm TeV} }
\def\diag{\mathop{\rm diag}\nolimits}
\def\Spin{\mathop{\rm Spin}}
\def\SO{\mathop{\rm SO}}
\def\O{\mathop{\rm O}}
\def\SU{\mathop{\rm SU}}
\def\U{\mathop{\rm U}}
\def\Sp{\mathop{\rm Sp}}
\def\SL{\mathop{\rm SL}}
\def\tr{\mathop{\rm tr}}

\def\IJMP{Int.~J.~Mod.~Phys. }
\def\MPL{Mod.~Phys.~Lett. }
\def\NP{Nucl.~Phys. }
\def\PL{Phys.~Lett. }
\def\PR{Phys.~Rev. }
\def\PRL{Phys.~Rev.~Lett. }
\def\PTP{Prog.~Theor.~Phys. }
\def\ZP{Z.~Phys. }


\baselineskip 0.7cm

\begin{titlepage}

\begin{flushright}
IPMU11-0048
\end{flushright}

\vskip 1.35cm
\begin{center}
{\large \bf Simplified R-Symmetry Breaking\\ and\\ Low-Scale Gauge Mediation
}
\vskip 1.2cm
Jason L. Evans$^{a}$, 
Masahiro Ibe$^{a,b}$, 
Matthew Sudano$^{a}$ and Tsutomu T. Yanagida$^{a}$
\vskip 0.4cm

${}^{a}${\it  Institute for the Physics and Mathematics of the Universe, \\
University of Tokyo, Chiba 277-8583, Japan}\\
${}^{b}${\it Institute for Cosmic Ray Research, University of Tokyo, Chiba 277-8582, Japan }

\vskip 1.5cm

\abstract{We argue that some of the difficulties in constructing realistic models of low-scale gauge mediation are artifacts of the narrow set of models that have been studied.  In particular, much attention has been payed to the scenario in which the Goldstino superfield in an O'Raifeartaigh model is responsible for both supersymmetry breaking and R-symmetry breaking.  In such models, the competing problems of generating sufficiently massive gauginos while preserving an acceptably light gravitino can be quite challenging.  We show that by sharing the burdens of breaking supersymmetry and R-symmetry with a second field, these problems are easily solved even within the O'Raifeartaigh framework.  We present explicit models realizing minimal gauge mediation with a gravitino mass in the eV range that are both calculable and falsifiable.
}
\end{center}
\end{titlepage}

\setcounter{page}{2}

\section{Introduction}

Much of the research on gauge mediation \cite{GM} has had little need for more than an effective description of the high-scale dynamics.  With the introduction of General Gauge Mediation\,\cite{Meade:2008wd} and associated works\,\cite{ggmb}, the low-scale parameterization of this class of models is quite robust.  Attempts at dynamically generating such effective descriptions, however, have had largely discouraging results.
Without explicit models the plausibility of these scenarios can be called into question. And, of course, if evidence consistent with gauge mediation is found at the LHC, we will have ample motivation to go beyond a spurion analysis.  It is in that spirit that we revisit the problem of constructing a fully consistent Minimal Gauge Mediation.

Since the relevant interactions in SUSY phenomenology are essentially fixed and have been thoroughly studied, SUSY model building has effectively been reduced to the problem of obtaining a spectrum within known bounds.  Let's discuss the players in turn.

\begin{itemize}

\item {\it Scalars} --
Scalar masses are famously easy to generate.  Indeed, one of the primary motivations for low-scale SUSY breaking is to keep a scalar (the Higgs) light.  To get sufficiently large scalar masses, we simply need to break supersymmetry at a sufficiently high scale.%
\footnote{The Higgs, of course, is a special case.  We won't discuss the Higgs sector in detail in this work, but the models we will discuss can be extended to include a $\mu_H$ and $B_\mu$ generating sector as in\,\cite{Evans:2010ru,Evans:2010kd}.}
In minimal gauge mediation, the scalars of the supersymmetric Standard Model first get masses at two-loops by coupling through gauge interactions to messenger fields.  These masses are schematically given by
\begin{equation}
m_0^2\sim\left(\frac{g^2}{16\pi}\right)^2\left|\frac{F}{M}\right|^2+{\cal O}\left(\left|F^4/M^6\right|\right),
\end{equation}
where $g$ is a Standard Model gauge coupling and $F$ gives the splitting of the messenger mass from its supersymmetric value, $M$.

Applying the experimental lower bounds on sparticle masses, we learn that $|F/M|\gtrsim 10-100$ TeV.  This is not strictly sufficient, however, because there are also constraints on relative magnitudes of sfermion masses from flavor changing neutral current data.  The flavor universality of gauge interactions gives adequate flavor degeneracy in gauge mediation to evade these constraints.  One must be mindful, however, of gravity-mediated effects that can reintroduce a flavor problem if the SUSY breaking scale is too high.

\item {\it Gravitino} --
In gauge mediation, though gravitational effects are otherwise decoupled, the gravitino plays a prominent role.  Since its mass is set by the reduced Planck scale, $M_{PL}$,
rather than the messenger scale,
\begin{equation}
m_{3/2}\sim\frac F{M_{PL}}
\end{equation}
it is generically the LSP and thus, assuming R-parity, it is stable.  Its stability affords it a conspicuous role in cosmology.  It is automatically a dark matter candidate,
and interestingly, the thermal relic abundance for $m_{3/2}\simeq 100$\,eV
is consistent with the observed dark matter density.
Unfortunately, however, such a light gravitino is not cold dark matter but corresponds to hot
dark matter and conflicts with structure formation data\,\cite{Viel:2005qj}.
The thermal relic density of the heavier gravitino which can be cold dark matter is, on the other hand,
much higher than  the observed density of dark matter.
If we allow an additional source of dark matter, however, a very light gravitino, $m_{3/2}\lesssim16$ eV, evades all constraints\,\cite{Viel:2005qj}.
Lowering the scale of SUSY breaking to this degree, however, is not easy.
We will discuss model building implications of such a light gravitino in section\,\ref{sec:lightgravitino}.

\item{\it Gauginos} --
In minimal gauge mediation, the gauginos present no difficulties as one finds schematically
\begin{equation}
m_{1/2}\sim m_0\sim\frac{g^2}{16\pi}\frac{F}{M}+{\cal O}\left(\left|F^2/M^3\right|\right).
\end{equation}
Realizing such gaugino masses, however, is a notorious problem in explicit models of gauge mediation.  The problem begins with the R-symmetry conflict: An unbroken R-symmetry, forbids gaugino masses; without the presence of such a symmetry, however, SUSY breaking is somewhat difficult \cite{Nelson:1993nf}. Of course, we can have it both ways. In theories with a spontaneously broken R-symmetry, SUSY breaking is generic and gaugino masses are generated, but two problems have been encountered.  First, breaking the R-symmetry is not trivial in practice \cite{Shih:2007av}.  Second, even when the symmetry is broken, many have found that the gaugino masses vanish to leading order in SUSY breaking \cite{Polchinski:1982an, inty, KS}.  We will discuss these issues further in section\,\ref{sec:Rbreaking}.

\item{\it R-axion} --
As for any global symmetry, the spontaneous breaking of an R-symmetry results in a massless particle.
It is, however, assumed that this symmetry is not present in a gravitational theory.
One may further introduce explicit and small breaking of the R-symmetry in the low-energy effective theory. This is tantamount to accepting the metastability of our vacuum \cite{Intriligator:2007py, KS, Giveon:2009yu}. By the explicit R-symmetry breaking effects in the supergravity,
this axion can be made heavy enough in some models\,\cite{Bagger:1994hh},
but its existence remains an imminently falsifiable prediction in many scenarios including those to be discussed here. In section\,\ref{sec:Raxion} we will elaborate on these points.
\end{itemize}

In the remainder of this paper we will present a framework in which each of the above challenges is met.  The low-energy effective action will be minimal gauge mediation after integrating out a SUSY-breaking sector along the lines of cascade gauge mediation (to be reviewed in section\,\ref{sec:lightgravitino}).  The key to addressing the gravitino problems, and the novelty in our approach, is that our messenger masses are generated at one-loop, so we can have a truly low-scale gauge mediation and a light gravitino.  The classic problems with breaking R-symmetry and generating gaugino masses simply aren't present in this framework.  As we will discuss in detail in the section\,\ref{sec:Rbreaking}, R-symmetry breaking is loop-induced and not strictly along the Goldstino pseudomodulus direction, so the analyses of \cite{Shih:2007av,KS} do not apply.  This method of SUSY and R-symmetry breaking is applied to an explicit class of models in section\,\ref{sec:appl}, where we discuss a dynamical embedding of the models.  This is needed to avoid a Landau pole problem.  In section\,\ref{sec:Raxion} we show that our R-axion decay constant is bounded on both sides and that future experiments should fully explore the currently allowed region.  The composition of the dark matter is not predicted.  Prospective candidates are discussed in section\,\ref{sec:conc} with our concluding remarks. In appendices we show that our mechanism of R-symmetry breaking can be sourced by D-term SUSY breaking, and we prove that in the F-term breaking case, we must have a field with R-charge other than 0 or 2.

\section{Attempts at Models with a Light Gravitino}\label{sec:lightgravitino}

The fundamental SUSY breaking scale can be expressed in terms of the gravitino mass as follows.
\begin{eqnarray}
\label{eq:Fgravitino}
\sqrt{F} \simeq (\sqrt 3 m_{3/2} M_{PL})^{1/2} \simeq
65\,{\rm TeV} \times \left( \frac{m_{3/2}}{1\,{\rm eV}}\right)^{1/2}\ .
\end{eqnarray}

In minimal gauge mediation, the soft supersymmetry breaking masses
of the superparticles in the SSM are roughly given by,
\begin{eqnarray}
\label{eq:soft}
m_{\rm soft} \simeq 10^{-2} \frac{F_S}{M}\gtrsim{\cal O}(10^{2-3})\,{\rm GeV}\ .
\end{eqnarray}
Here, $M$ is the messenger scale and $F_S$ is the mass splitting of the messenger multiplets.
These mass scales satisfy $F_S<M^2$  and $F_S < F$,
where the first condition ensures that our messengers are non-tachyonic and the second one
says that the mass splitting is smaller than the total size of supersymmetry breaking.
By combining these conditions with \eqrefc{eq:Fgravitino} and \eqrefc{eq:soft}, we find that having an eV scale gravitino while having sufficiently heavy superpartner requires
\begin{eqnarray}
\label{eq:lowscale}
M \sim \sqrt{F_S} \sim \sqrt{F} \sim {\cal O}(10-100)\,{\rm TeV}\ .
\end{eqnarray}

For the SUSY breaking mass splitting in the messenger sector to be comparable
to the fundamental SUSY breaking scale, i.e. $F_S/F={\cal O}(1)$, we need to have
the fundamental SUSY breaking sector and the messenger sector
couple via interactions that are not too suppressed. As we briefly discuss below, this fact makes building a model with a very light gravitino rather difficult.

\subsubsection*{Tree-level Interactions}

The simplest idea for realizing models with $F_S\sim F$ is to connect the messenger
and the SUSY breaking sectors by tree-level interactions in the superpotential
with coupling constants of order one.
For example, if we require that the SUSY breaking vacuum is the absolute minimum of the model,
the model can be realized by an O'Raifeartaigh model with
the superpotential,
\begin{eqnarray}
\label{eq:tree}
W = \mu^2 S + (m_{ij} + \l_{ij} S) \bar{\psi}_i \psi_j\ .
\end{eqnarray}
Here, $S$ represents a SUSY breaking field,  $\psi$, $\bar\psi$ are messenger fields,
$\mu$ and $m_{ij}$ are mass parameters, and the $\l_{ij}$ are coupling constants.
In this class of models, the SUSY breaking vacuum is the absolute minimum
when the model possesses an R-symmetry under which $S$
has charge $2$ for $\det m\neq 0$\,\cite{KS} and $\l m^{-1}\l = 0$\,\cite{Sato:2009dk}.%
\footnote{
Notice that the R-symmetry is also a necessary condition
for the models not to have SUSY vacua when the superpotential
is generic under the symmetries\,\cite{Nelson:1993nf}.
If the models are restricted to have renormalizable interactions,
the necessary condition can be relaxed to a discrete R-symmetry.}

In this class of models\,\cite{inty} (and a broader class, as shown in \cite{KS}), the leading contribution to
the gaugino masses appears only at order $F_S^3$ (first calculated in \cite{inty}).
This is opposed to the naive expectation in \eqrefc{eq:soft},%
\footnote{
We are assuming that the R-symmetry is spontaneously broken
by the expectation value of the SUSY breaking field, i.e. $\vev S =O(M)$,
which is the optimal choice for generating large gaugino masses.}
\begin{eqnarray}
\label{eq:gaugino0}
 m_{\rm gaugino} \sim  \frac{\a}{4\pi} \frac{F_S}{M} \left| \frac{F_S}{M^2}\right|^2\ .
\end{eqnarray}
For the parameters satisfying \eqrefc{eq:lowscale}, the factor $|F_S/M^2|^2$
may appear to be of order unity.
Detailed numerical analysis, however, has shown that the predicted gaugino masses
are rather suppressed and have been almost excluded by Tevatron constraints
on the neutralino/chargino masses
for $m_{3/2}\lesssim 16$\,eV\,\cite{Ibe:2005xc,Sato:2009dk,Sato:2010tz}.

Larger gaugino masses are possible in models
based on tree-level interactions by allowing an instability.
For example, naively we expect gaugino masses like those in \eqrefc{eq:soft} are generated
by the superpotential
\begin{eqnarray}
\label{eq:tree2}
W = \mu^2 S + (m + \l S) \bar{\psi}\psi\ ,
\end{eqnarray}
which breaks R-symmetry explicitly.
The model has a SUSY preserving vacuum at $\vev{\bar\psi \psi} = -\mu^2/\l$.
In this class of models, to ensure sufficiently long lifetime of the SUSY breaking vacuum
one must increase the messenger mass
$m$, which suppresses superparticle masses for a given gravitino mass.
For $m_{3/2}\lesssim 16$\,eV, the numerical analysis has shown that the upper bound on the gluino/squark masses is about 1\,TeV\,\cite{Hisano:2008sy}.
Besides, this class of models requires some justification
for the SUSY breaking vacuum to be chosen
in the evolution of the universe,
regardless of its higher energy compared with
the energies of the nearby supersymmetric vacua.

\subsubsection*{Direct/Indirect Gauge Mediation}

Another approach that may give $F_S \sim F$ is to assume that
the messenger fields are charged under the gauge symmetry
which is responsible for dynamical SUSY breaking (DSB).
Such models are often called direct or semi-direct gauge mediation models.
In direct mediation models, the messengers play a role in SUSY breaking
dynamics, while they do not in semi-direct mediation models.
(A precise definition of direct gauge mediation is given, for example, in
\cite{Dine:2007dz,Carpenter:2008wi}.)
In these classes of models, since messengers strongly couple to the SUSY breaking fields,
the size of the messenger mass splittings are expected to be comparable to the fundamental SUSY breaking scale.

In the direct mediation models, however, the need for a large flavor symmetry in the DSB sector
usually implies that the DSB gauge group is rather large.
Thus, to avoid Landau poles of the Standard Model gauge interactions below the GUT scale,
both the messenger and SUSY breaking scales are pushed up,
which precludes the possibility of having a light gravitino.%
\footnote{The Landau pole problems may be ameliorated
if the messenger fields receive large positive anomalous
dimensions under the renormalization group evolution
between the GUT and the messenger scales\,\cite{Ibe:2007wp}.
}

The messenger and SUSY breaking scales can be much lower in semi-direct gauge mediation models\,\cite{Ibe:2007wp,Seiberg:2008qj,Elvang:2009gk}
(see also \cite{Randall:1996zi,Izawa:2005yf} for  earlier attempts),
where the gauge group for the DSB can be smaller.
In this class of models, however, the leading contribution to the gaugino mass as in \eqrefc{eq:soft}
is again vanishing (see \cite{Izawa:2008ef} for details).
Therefore, both direct and semi-direct gauge mediation models have difficulty producing a gravitino mass in the eV range.

\subsubsection*{Cascade Gauge Mediation}

In \cite{Ibe:2010jb}, a class of models termed cascade gauge mediation
was proposed with the following features.
\begin{itemize}
\item Gravitino mass in the eV range
\item Sufficiently large gaugino masses
\item A SUSY-breaking vacuum with a long enough lifetime
\item Perturbativity up to the GUT scale
\end{itemize}
Since our construction will mirror that of cascade gauge mediation in some ways, we will describe it in some detail.

The mechanism employs some of the tools proposed in previous gauge mediation
models\,\cite{Dine:1993yw,Dine:1994vc,Dine:1995ag}.  Namely two SUSY-breaking fields are introduced, one of which is an ordinary tree-level SUSY-breaking field.  The other field is the one that couples to the messengers in the superpotential and it only breaks SUSY and only couples to the ``fundamental'' SUSY-breaking field through radiative corrections to the K\"ahler potential.

The simplest such model is described by the following K\"ahler potential and superpotential.
\begin{eqnarray}
\label{eq:CCM}
 K &=& |Z|^2 + |S|^2 + \frac{c}{\L^2} |Z|^2 |S|^2 +\cdots\ ,\cr
 W &=& \mu^2 Z + k S \widetilde\Phi \Phi + \frac{h}{3}S^3\ ,
\end{eqnarray}
where $\mu$ and $\L$ are dimensionful parameters, and $k, h$ are dimensionless coupling constants.
\footnote{Here, we may assume that $h$ and $k$ are positive without losing genericity.}
Here, $Z$ is the fundamental SUSY-breaking field, while $S$ now represents the
secondary SUSY breaking field.
Note that the messenger sector possesses a supersymmetric vacuum in the limit $c\to 0$.

Once the primary sector breaks SUSY, i.e. $|F_Z|  = \mu^2$, the linking term in the K\"ahler potential induces a
SUSY breaking soft mass for $S$, and the scalar potential of the secondary SUSY-breaking
field is given by,
\begin{equation}
V(S) \simeq  m_{S}^2 |S|^2 + | h S^2|^2,\qquad m_S^2 = -c \frac{|F_Z|^2}{\L^2}\ .
\end{equation}
For $m_S^2 < 0$, the secondary SUSY-breaking field obtains a non-vanishing expectation value,%
\footnote{
For generic models of cascade gauge mediation, the secondary sector breaks supersymmetry
even for $m_S^2 >0$\,\cite{Ibe:2010jb}.
}
\begin{eqnarray}
\label{eq:vevSapp}
 \vev S \simeq \frac{|m_S|}{\sqrt 2 h}\ .
\end{eqnarray}
which breaks SUSY by,
\begin{eqnarray}
\label{eq:vevFapp}
 F_S =  h\vev S^{*2}\ \simeq  \frac{|m_S^2|}{2 h}\ .
\end{eqnarray}
Thus, in this model, secondary SUSY breaking
is initiated by spontaneous R-symmetry breaking which is, in turn, triggered by
fundamental SUSY breaking.

Since the structure of the messenger and the secondary fields coupling is nothing but that of minimal gauge mediation, the gaugino masses are given by
\begin{eqnarray}
\label{eq:gaugino}
 m_{a} = \frac{\alpha_a}{4\pi}\frac{F_S}{\vev S} = \frac{\alpha_a}{4\pi} h \vev S \simeq \frac{\alpha_a}{4\pi}\frac{|m_S|}{\sqrt2}\ .
\end{eqnarray}
Thus, for models with $m_S \sim \sqrt{F_Z}$ which corresponds to
$\L\sim \sqrt{F_Z}$ and $c =O(1)$,
the conditions in \eqrefc{eq:lowscale} for a very light gravitino are satisfied
while having heavy gauginos.

To realize models with $\L\sim \sqrt{F_Z}$ and $c =O(1)$ is, however, not an easy task.
In previous cascade gauge mediation models\,\cite{Dine:1993yw,Dine:1994vc,Dine:1995ag}, for example,
the connecting term in \eqrefc{eq:CCM} is generated at three loops, which leads to $\L\gg \sqrt{F_Z}$ and/or $|c|\ll 1$.

In the model given in \cite{Ibe:2010jb},
the connection term is also generated at higher-loop level.
In that model, however, all the interactions are expected to be very strong,
so that the conditions in \eqrefc{eq:lowscale} are realized, and hence,
the model allows the gravitino mass to be in the eV range.
The drawback of the model in \cite{Ibe:2010jb} is that
the model cannot predict the sign of the sfermion masses squared
due to the non-calculable contribution of the strong interactions.

Therefore, the framework of cascade gauge mediation
is a promising way for realizing a gravitino mass in the eV range. However,
it is desirable to construct models where $m_S^2$ is generated at one-loop,
so that the conditions in \eqrefc{eq:lowscale} can be satisfied in a perturbative framework.

\section{Non-Goldstino R-symmetry Breaking at One-loop}\label{sec:Rbreaking}

In this section, we discuss the mechanism of R-symmetry breaking that we utilize in our models.

In \cite{Shih:2007av}, the problem of R-symmetry breaking was considered in the set of O'Raifeartaigh models defined by the superpotential
\begin{equation}\label{shihw}
W=fX+\frac12(M^{ij}+XN^{ij})\phi_i\phi_j+\dots,\qquad\vev{\phi_i}=0\ .
\end{equation}
It was shown that one-loop spontaneous R-symmetry breaking
is only possible when some fields are assigned R-charge other than $0, 2$.
The proof made use of an expression for the one-loop mass of the charge-two pseudomodulus, $X$, in terms of a positive semi-definite and a negative semi-definite contribution.  It was shown that the latter contribution vanishes if every field can be assigned an R-charge of $0$ or $2$. This illuminated the observed absence of R-symmetry breaking in ungauged O'Raifeartaigh models.

One limitation of this analysis is that only models with a single pseudo-modulus (the superpartner of the Goldstino) are allowed.  The important difference in our approach is that we allow a second pseudo-modulus and look for R-symmetry breaking in a direction other than the Goldstino pseudo-modulus direction. Although these models are not included in \eqrefc{shihw}
we still find that a field with R-charge other than $0$ or $2$ is required
(see appendix\,\ref{sec:NCR}).%

\subsubsection*{R-Symmetry Breaking in O'Raifeartaigh models}

Here, we discuss models with calculable R-symmetry breaking that is radiatively induced by SUSY breaking. We restrict ourselves to models where the R-charge of our non-Goldstino pseudomodulus is undetermined. We will also restrict ourselves to models with a global SUSY breaking minimum. A broad class%
\footnote{This class has been chosen to have a strictly negative $m_S^2$.  More general models with positive contributions that are subdominant to the the negative ones are not difficult to construct. Adding interactions of the form, $S B_iB_j$, can generate a tachyonic mass for $S$, however, they tend to require fine tuning.}
of models with these features is
\begin{equation}
W=Z(\mu^2+g_{ij}B_iB_j)+M_{ij}B_iC_j+M'_{ij}B_iD_j+\lambda_{ij}SD_iE_j+M''_{ij}X_iY_j
\label{GeMs} \ .
\end{equation}
where the $X_i$ and $Y_i$ may be any fields other than $S$, $Z$, or the $C_i$. In any model of this form, $S$ will have a tachyonic mass at one loop.  This can be seen by tracing the effects of SUSY breaking.  $S$ is chosen to be massless at tree-level, so the one-loop mass must vanish in the limit of unbroken supersymmetry by the perturbative non-renormalization of the superpotential.
When SUSY is broken, the masses of the bosons are split, but the fermion masses stay the same.
This means that at one loop, the only negative contributions to $m_S^2$ that are affected by SUSY breaking are the boson loops coming from the trilinear interactions,
 \begin{eqnarray}
{\cal L} =M'_{ij}B_j\lambda_{ik}^*S^*E_k^*+h.c.
\end{eqnarray}
In the above model, the only fields that get split masses at tree level are the $B_i$. Since these fields only interact with $S$ through trilinears, only the negative contribution to $m_S^2$ is enhanced by SUSY breaking. Naturally, $m_S^2$ is negative.

Another important feature of this class of models is that the SUSY breaking minimum is a global minimum as long as $\det M\ne 0$. In this case, the only solution to
\begin{equation}
-F_{C_i}^*=M_{ij}B_{j}=0
\end{equation}
is $B_i=0$. If $B_i=0$ for all $i$ then $|F_Z|=\mu^2$ at the global minimum of the potential and SUSY is broken%
\footnote{Depending on the mass parameters, $F_Z=\mu^2$ may not be the global minimum.  However, the global minimum of these models is always a SUSY breaking minimum.}.

Now that we have found a class of models with $m_S^2<0$,
we will consider the simplest case,
\begin{equation}
W= Z(\mu^2-gB^2)+m(BC+BD)+\lambda S D E +m' (EF+GD) \ .
\label{eq:Sup1}
\end{equation}
We have kept two mass scales $m$ and $m'$ for later use.
This model has an R-symmetry (under which $R[S]=-R[E]=R[F]-2$ is undetermined),
and a $Z_2$ symmetry under which all the superfields are odd except
for $Z$ and $S$.
Under these symmetries, the terms
\begin{eqnarray}
\label{eq:NGterms}
GC\ , \quad SCE \ ,
\end{eqnarray}
are allowed.  Including them would introduce unwanted supersymmetric vacua.
Those problematic terms, however, can be naturally suppressed by assuming
that there is an additional $U(1)$ symmetry which has been spontaneously broken
at around the Planck scale only by a positively charged suprion field $\phi_+$ (charge $+1$).
By assuming that $G$ and $E$ have charge $+1$, $D$ and $F$ have charge $-1$,
and the other fields have charge $0$,
the terms in \eqrefc{eq:NGterms} are suppressed under the broken symmetries
with the help of the holomorphic nature of the superpotential,
while the terms appearing in \eqrefc{eq:Sup1} are not suppressed as long as
$\vev{\phi_+}=O(M_{PL})$.
In this sense, the model in \eqrefc{eq:Sup1} is generic under the (broken) symmetries.

Of course, the term $hS^3/3$ is also allowed.  We will include this term when we consider cascade gauge mediation. The cubic term of $S$ has no effect on the one-loop mass, but it fixes $R[S]=2/3$, making the vev of this field R-symmetry breaking.

\begin{figure}[tbp]
\begin{center}
\includegraphics[width=0.6\textwidth]{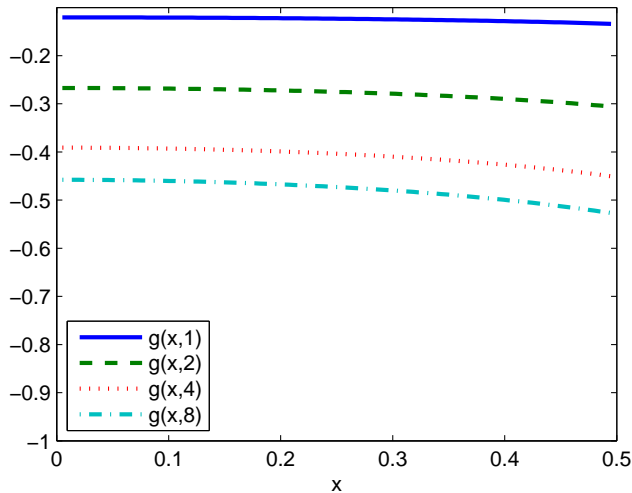}\label{Graph1}
\end{center}
\caption{The function $g(x,y)$ ($x=g\mu^2/m^2, y=m^\prime/m$)
given in \eqrefc{eq:massSQ}.
The figure shows that the mass squared of $S$ is always negative in these models,
which leads to spontaneous R-symmetry breaking. $g(x,y)$ is cut off at $x=1/2$ because of the emergence of a tachyonic mass.
}
\label{fig:fig1}
\end{figure}
The one-loop squared mass of $S$ in this model is quite complicated but takes the form
\begin{eqnarray}
\label{eq:massSQ}
 m_S^2 = \frac{g^2\lambda^2}{16\pi^2}\frac{\mu^4}{m^2}g(x,y)\ ,
 \end{eqnarray}
 where $x=\mu^2/m^2$ and $y=m'/m$. Since $g(x,y)$ is only weakly dependent on $x$, we only need its leading order contribution
 \begin{equation}
 g(x,y)\simeq g(0,y)  =\frac{12y^2(y^2+2)}{(y^4+4)^{5/2}}\left[ \ln\left(\frac{2+y+\sqrt{4+y^4}}{2+y-\sqrt{4+y^4}}\right)-\frac{\sqrt{4+y^4}
 \left(y^4+12y^2+4\right)}{6y^2(y^2+2)}\right].
 \label{OrMo}
\end{equation}

To supplement our approximation, we show the $x$ dependence of $g(x,y)$ for a few values of $y$ in Figure\,\ref{fig:fig1}.
From the figure it is apparent that $g(x,y)$ is negative for all values of $x, y$ and is
typically in the range $g(x,y)\simeq -(0.2-0.5)$.  Clearly we find that $R$ symmetry is successfully broken at one loop and
\begin{eqnarray}
 \vev S \simeq \frac{g \l}{4\pi h}\frac{\mu^2}{m}\ .
\end{eqnarray}

\section{Gravitino Mass in the eV Range}\label{sec:appl}
Now let us apply the models in section\,\ref{sec:Rbreaking}
to cascade gauge mediation, and examine whether the models
allow a gravitino mass in the eV range.
For that purpose, we add the cubic term of $S$ and messenger coupling
to the model in \eqrefc{eq:Sup1};
\begin{equation}
W= Z(\mu^2-gB^2)+m(BC+BD)+\lambda S D E +\widetilde m (EF+GD) + \frac{h}{3} S^3 + k S \psi\bar{\psi}\ .
\label{eq:Sup2}
\end{equation}
By integrating out everything except for $S$ and the messengers, this model is reduced to cascade gauge mediation,
\begin{eqnarray}
 W  =  \frac{h}{3} S^3 + k S \psi\bar{\psi}\ ,
\end{eqnarray}
and
\begin{eqnarray}
 V_{\rm soft}  \simeq  m_S^2 |S|^2\ ,
\end{eqnarray}
where $m_S^2<0$ is given in \eqrefc{eq:massSQ}.

In order to have the largest possible gaugino masses for $\mu \lesssim 260$\,TeV
(i.e. $m_{3/2}<16$\,eV, see \eqrefc{eq:Fgravitino}),
the coupling constants $g$ and $\lambda$ should be as large as possible
and $m$ should be as small as possible.
From our numerical analysis, however, we found that to avoid tachyonic modes, we must have
\begin{eqnarray}
  g \mu^2 \lesssim 0.5\ m^2\ ,
\end{eqnarray}
for $k = {\cal O}(1)$.

With this constraint, the soft mass of $S$, is bounded from above,
\begin{eqnarray}
\label{eq:mSup}
  |m_S| \simeq \left(\frac{g^2\lambda^2}{16\pi^2}\frac{\mu^4}{m^2}|g(x,y)| \right)^{1/2}
  \lesssim \frac{g^{1/2}\lambda}{8\pi}\mu\ ,
\end{eqnarray}
where we have used $g(x,y)\simeq -0.5$ in the final expression.
In cascade gauge mediation, the bound on $m_S^2$ results in a bound on the gaugino masses. For example, we obtain,
\begin{eqnarray}
 m_{\rm gluino} \lesssim 50\,{\rm GeV} \times N_{\rm mess}g^{1/2}\lambda
 \left(\frac{\mu}{260\,{\rm TeV}}\right)\ ,
\end{eqnarray}
where $N_{\rm mess}\leq 5$ is the number of messengers (see \eqrefc{eq:gaugino}).%
\footnote{
Here, we have approximated that the pole mass of the gluino
is roughly enhanced by $25$\% from the value of $m_{3}$
at the weak scale.
For the model with a large hierarchy between the gaugino masses
and sfermion masses, the pole mass enhancement can be slightly
larger.
Even with such an enhancement, it is unlikely that the gluino mass exceeds
the lower mass bound at the LHC experiment.
}
Therefore, the gluino mass does not satisfy
the current lower limit placed by the ATLAS collaboration\,\cite{Collaboration:2011hh}%
\footnote{See also the result from the CMS collaboration\,\cite{Khachatryan:2011tk}.}
$m_{\rm gluino}\gtrsim 700$\,GeV\,
as long as $g^{1/2} \lambda \lesssim 1$ even for $N_{\rm mess}=5$.

Notice that we cannot freely take large values of $g$ and $\lambda$.
For example, if we require that both $g$ and $\lambda$ are perturbative up to
some high energy scale (such as the GUT scale), they are expected to be at most below 1 or so.%
\footnote{
By assuming that $h$ is much smaller than $\l$ and $g$, the renormalization group equations
are solved by
\begin{eqnarray}
 \l^2(\mu_R) \simeq \frac{\l^2(M_{\rm GUT})}{1-\frac{3\l^2(M_{\rm GUT})}{8\pi^2}\ln(\mu_R/M_{\rm GUT})}\ , \quad
  g^2(\mu_R) \simeq \frac{g^2(M_{\rm GUT})}{1-\frac{10g^2(M_{\rm GUT})}{8\pi^2}\ln(\mu_R/M_{\rm GUT})}\ ,
\end{eqnarray}
which give $\l(10^5\,{\rm GeV})\simeq 1$ and $g(10^5\,{\rm GeV})\simeq 0.6$ for
 $\l(M_{\rm GUT})=g(M_{\rm GUT})\simeq 4\pi$.
 }
Therefore, to have an acceptably heavy gluino, we need to allow $g$ or $\lambda$
 to blow up below the GUT scale, which in turn requires
some other description of the model, i.e. a UV completion.\footnote{Alternate solutions for such Landau pole problems have been proposed.  See, for example, \cite{Abel:2008tx}.}

Fortunately, we find that the model in \eqrefc{eq:Sup2} can be embedded
into a dynamical SUSY breaking model based on a vector-like $SU(2)$ gauge theory
developed in \cite{Izawa:1996pk,Intriligator:1996pu}.
There, the coupling constant $g$ in the effective description in \eqrefc{eq:Sup2}
of the dynamical SUSY breaking model can be rather large, and the predicted
gluino mass can be above the current experimental limit.
In the rest of this section, we discuss how the above model in \eqrefc{eq:Sup2}
is embedded in this dynamical SUSY breaking model.

\subsection{Embedding in Dynamical SUSY Breaking Model}
Now let us begin with a brief review of the dynamical SUSY breaking model
based on $SU(2)$ gauge theory\,\cite{Izawa:1996pk,Intriligator:1996pu}.
The model consists of four fundamentals $Q_i(i=1\cdots 4)$ and six singlets,
$Z_{ij}=-Z_{ji} (i,j=1\cdots 4)$.
The $Q$'s and $Z$'s couple in the classical superpotential,
\begin{eqnarray}
\label{eq:SQQ}
 W &=& g_{ij}^{kl} Z_{ij}Q_{k}Q_{l}\ , \, (i<j)\ ,\cr
      &=& g_0 Z_0 (QQ)_0 + g^\prime Z_{a} (QQ)_a\ ,  (a=1\cdots 5)\ ,
\end{eqnarray}
where the $g$'s are coupling constants.
The maximal global symmetry this model may have is
$SU(4)\simeq SO(6)$ symmetry which requires $g_{ij}^{kl} = g$.
For simplicity, in the second expression we have rearranged the fields and
coupling constants assuming there exist an $SO(5)\subset SO(6)$ global symmetry.
The subscript $a=1\cdots 5$ corresponds to a fundamental representation
of $SO(5)$.
In this model,  SUSY is broken dynamically  due to the tension
between the $F$-term conditions of $Z$'s and
the quantum modified constraint\,\cite{Seiberg:1994bz} ${\rm Pf}(Q_{i}Q_{j}) =\L_{\rm dyn}^{2}$.

Below the dynamical scale $\Lambda_{\rm dyn}$, the light degrees of freedom are the composite operators $M_{A}=(QQ)_A$
and $Z_{A}$, ($A=0,1\cdots 5$).
In terms of the composite operators, the quantum moduli constraint is given by,
\begin{eqnarray}
\label{eq:M0}
M_AM_A =\L_{\rm dyn}^2\ .
\end{eqnarray}
Here, we have assumed that
the effective composite operators $M_A$ are canonically normalized.

By assuming that $g_0$'s are perturbative and $g_0 < g^\prime$,
we may parametrize the deformed moduli space by,
\begin{eqnarray}
  M_0 = \sqrt{  \L_{\rm dyn}^2 - M_{a}M_{a}}\,.
\end{eqnarray}
By plugging it into the effective superpotential in \eqrefc{eq:SQQ},
we obtain the effective O'Raifeartaigh model,
\begin{eqnarray}
\label{eq:Weff1}
W_{\rm eff} &\simeq& g_0 \,\L_{\rm dyn}^{2} Z_{0}
- \frac{g_0}{2}\, Z_{0}M_{a}M_{a}
+g'\,\L_{\rm dyn}Z_{a}M_{a} + O(M_{a}^{4})\ .
\end{eqnarray}
In this effective description, SUSY is broken by,
\begin{eqnarray}
\label{eq:vac}
 F_{Z_{0}} = g_0 \L_{\rm dyn}^{2}\ .
\end{eqnarray}

Now we can relate the dynamical SUSY breaking model
to the model discussed in the previous section by matching the fields and parameters;
\begin{eqnarray}
 Z_0 &\to& Z\ ,\quad  M_a \to B_a\ , \quad Z_a\to D_a\ , \cr
 g_0\L_{\rm dyn}^2 &\to& \mu^2\, \quad g^\prime \L_{\rm dyn} \to m\ ,
 \quad g_0 \to 2 g\ .
\end{eqnarray}
With this dictionary, the above effective O'Raifeartaigh model is rewritten as
\begin{eqnarray}
\label{eq:Weff2}
W_{\rm eff} &\simeq& Z\left(\mu^2 - g B_{a}B_{a}\right)+ m B_{a}D_{a} \ ,
\end{eqnarray}
where we have neglected the higher dimension operators. 
Finally, let us introduce four $SO(5)$-vector superfields, $C_a$, $E_a$, $F_a$ and $G_a$,
a singlet field $S$ and pairs of messengers.
\begin{eqnarray}
\label{eq:Sup3}
W_{\rm eff} &\simeq& Z\left(\mu^2 - g   B_{a}^{2} \right)
+m B_{a}(D_{a} + C_a) + \widetilde m (F_a E_a + G_a D_a)\cr
&&+ \,\lambda S D_a E_a +\frac{h}{3} S^3 + k S \bar{\psi}\psi\ ,
\end{eqnarray}
where $SO(5)$ indices are contracted.
In this way, we can embed the previous R-symmetry breaking model into
a dynamical SUSY breaking model.

The important feature of this embedding is that the superfields $S$, $D_a$, $E_a$
are not the composite but elementary fields.
Thus, the trilinear couplings $SDE$ and $S^3$ are not suppressed in the UV theory.
That is, in the UV theory, the model in \eqrefc{eq:Sup3} is realized by,
\begin{eqnarray}
\label{eq:Sup4}
W_{\rm tree} &\simeq&
 g_0 Z_0 (QQ)_0 + g^\prime D_{a} (QQ)_a + g^\prime C_{a} (QQ)_a
 + \widetilde m (F_a E_a + G_aD_a)
\cr
&&+ \,\lambda S D_a E_a +\frac{h}{3} S^3 + k S \bar{\psi}\psi\ ,
\end{eqnarray}
which shows that $\lambda$, $h$ and $k$ are unsuppressed%
\footnote{By adding $\lambda' S'D_aJ_a+m_JJ_aK_a+\alpha'S'^3+k'S'H\bar{H}$ to the above superpotential, a $\mu_H/B_{\mu}$ term can be generated. If $\lambda'\ll \lambda$ and/or $m_J\gg m$, $\mu_H/B_{\mu}$ will have the same size as the soft masses.}.

The advantage of this embedding is that the effective coupling $g$,
which is related to $g_0$ in the UV theory, is not directly constrained by the
Landau pole problem.
In particular, the coupling $g$ (or $g_0$) is expected to become very large
around the dynamical scale $\L_{\rm dym}$ due to the renormalization
group effects of the $SU(2)$ gauge interactions, which enhance the
coupling $g$ (or $g_0$) in the lower energy.
Besides, since we have $SO(5)$ vectors,
 the one-loop contribution to $m_S^2$ is enhanced by a factor of $5$.
 With these two effects, the gluino mass upper bound is enhanced to
 \begin{eqnarray}
 \label{eq:gluinoOPT}
 m_{\rm gluino} \lesssim 2\,{\rm TeV} \times \lambda
 \left(\frac{N_{\rm mess}}{5}\right)
 \left(\frac{g}{4\pi}\right)^{1/2}
 \left(\frac{\mu}{260\,{\rm TeV}}\right)\ .
\end{eqnarray}
Here, we have taken $g\simeq 4\pi$ as the optimal choice.%
\footnote{Here, again, we have assumed that the pole mass of the gluino is roughly
enhanced by $25$\% compared with the value of $m_3$ at the weak scale.}
Therefore, by considering the dynamical SUSY breaking model behind the R-breaking model,
we find that the models with {\cal O}(eV) gravitino mass can be constructed with a heavy enough gluino, a stable vacuum, and perturbative unification.%
\footnote{
The above upper bound is quite optimistic since we assumed $g\simeq 4\pi$
where the perturbative analysis of $m_S^2$ is no longer reliable.
In such a non-perturbative region, however, the sign of the sfermion masses squared
of the SSM is not directly affected by the large coupling constant $g$ in this model.
This is a big difference from the previous models such as the one in \cite{Ibe:2010jb}
where the sign of the sfermion mass squared is incalculable in the non-perturbative region.
}
Furthermore, the predicted upper bound is well within the reach of the LHC experiment,
and hence, the class of models with the gravitino mass in the eV range based on
R-symmetry breaking at one loop will be ruled out or supported by data in near future.

\section{R-axion Properties}\label{sec:Raxion}

Before concluding this paper, let us discuss the important properties
of the R-axion in this class of models.
The R-axion is the pseudo-Nambu-Goldstone boson resulting from spontaneous R-symmetry
breaking\,\cite{Nelson:1993nf}.
The R-axion would be a true Nambu-Goldstone field if the R-symmetry were exact, but this is not the case. It is at least explicitly broken by the cosmological constant in supergravity, which yields a mass for the R-axion
 \cite{Bagger:1994hh}.

To see how the axion mass is generated, let us first study a simple example.  Consider the superpotential terms
\begin{eqnarray}
 W \supset \mu^2 Z + m_{3/2}M_{PL}^2\ ,
\end{eqnarray}
and assume that the R-charge-two field, $Z$, gets an expectation value providing the sole source of R-symmetry breaking in the limit $m_{3/2}\to0$.
\begin{eqnarray}
 Z =\vev{Z}e^{2i a/f_R}\ ,\qquad f_R = 2\sqrt 2\vev{Z} \ .
\end{eqnarray}
This defines the R-axion, $a$.
We can then compute the potential and see that the explicit R-symmetry breaking constant term provides a mass for $a$.
\begin{eqnarray}
\label{eq:Rbreaking1}
 V_{\mbox{\tiny R-breaking}} = -2 m_{3/2} \mu^2 Z + h.c.
\end{eqnarray}
Using $\mu^2 \simeq \sqrt{3}m_{3/2}M_{PL}$, we find
\begin{eqnarray}
  m_R^2 \simeq \frac{8}{\sqrt{2}} \frac{m_{3/2}\mu^2}{f_a}
  \simeq 8{\sqrt{\frac{3}{2}}} \frac{M_{PL}}{f_R} m_{3/2}^2
 \ .
\end{eqnarray}

In the models discussed in the previous section, the mechanism is somewhat different.
There the source of spontaneous R-symmetry breaking
is the vev of $S$, which couples to other fields through dimensionless interactions
at tree level (see \eqrefc{eq:Sup2}).
In this case, the explicit R-symmetry breaking term
analogous to
the one in \eqrefc{eq:Rbreaking1}
is induced from the higher-dimensional operators in the K\"ahler potential,
which can be expressed as the anomaly-mediated 
A-term\,\cite{Randall:1998uk,Giudice:1998xp},
\begin{eqnarray}
{\cal L} &=&\frac{h}{3}A_h S^3 +h.c. \cr
A_h &= & h \gamma_S \times m_{3/2}\ ,\cr
\g_S& = & 2 h^2 + 5 N_{\rm mess} k^2 + 5 \l^2\ .
\end{eqnarray}
The first contribution to the anomalous dimension of $S$, $\gamma_S$, 
comes from the $S^3$ interactions,
the second one is from the $S\bar{\psi}\psi$ interactions,
and the third one from the contributions of $S D^aE_a$ interactions.
The induced explicit R-breaking term is then given by,
\begin{eqnarray}
 V_{\mbox{\tiny R-breaking}} \simeq
 - \frac{1}{16\pi^2}
 \frac{h}{3}
 \left(
  2 h^2 + 5 k^2 N_{\rm mess}
  + 5\l^2 
 \right)
 m_{3/2}S^3  + h.c.
\end{eqnarray}
Thus, the R-axion defined by,
\begin{eqnarray}
S &=& \vev{S}e^{\frac{2i  a}{3f_R}}\ , \quad (f_R = 2 \sqrt{2}\vev{S}/3)\ ,\cr
F_S &=& \vev{F_S}e^{\frac{-4 ia}{3f_R}}\ ,
\end{eqnarray}
obtains the mass,
\begin{eqnarray}
m_R^2 \simeq \frac{9}{16\sqrt{2}\pi^2} \left(
  4 h^2 + 5 k^2 N_{\rm mess} + 5\l^2
 \right) m_{3/2} |m_S|\ .
\end{eqnarray}
In the final expression, we have used \eqrefc{eq:vevSapp}, and \eqrefc{eq:vevFapp}.
For $k=O(1)$, the contribution from the messenger loop dominates the R-axion mass and gives,
\begin{eqnarray}
m_R \lesssim 40\,{\rm MeV}\times k\l^{1/2}\left(\frac{N_{\rm mess}}{5}\right)^{1/2}
\left(
\frac{g}{4\pi}
\right)^{1/4}
\left(
\frac{m_{3/2}}{16\,\rm eV}
\right)^{3/4}\ ,
\end{eqnarray}
where we have used the upper bound on $m_S$ in \eqrefc{eq:mSup}.\footnote{We multiply the upper bound by $5^{1/2}$ with the embedding
into the dynamical model in mind.} We should note that the fact that our R-symmetry is anomaly free---even in our dynamical completion---is important.  As noted in \cite{Abel:2007jx} there is an enhancement to the R-axion mass when the R-symmetry is anomalous.

The decay constant $f_R$ of the R-axion with a mass of tens of MeV is constrained
by the contribution to the invisible decay mode of the $K$-meson, $K^+\to \pi^+ + a$,
which requires $f_R \gtrsim 2\times 10^5$\,GeV.
The decay constant is also constrained
by the supernova SN 1987a, which requires $f_R \lesssim 10^6$\,GeV.
Thus, by remembering that $|m_S| \lesssim 10^5$\,GeV for $m_{3/2}\lesssim 16$\,eV,
we find that the coupling constant $h$ needs to satisfy $h\lesssim 0.1$, so that $f_R$
is in between these constraints.
Furthermore, future neutrino experiments are expected to discover
the R-axion in this range or close the gap between these constraints\,\cite{Essig:2010gu}.

\section{Conclusions and Unresolved Issues}\label{sec:conc}

In this paper, we have discussed challenges in building explicit realistic models of gauge mediation. Focussing on the very light gravitino scenario, we found that in the framework of cascade gauge mediation we can construct viable models.  The key was a one-loop breaking of R-symmetry that was not aligned with the goldstino.

As is well known, it is quite challenging to find perturbative models with a very light gravitino, sufficiently heavy gaugino masses, and a stable SUSY-breaking vacuum.
Examining models based on one-loop R-symmetry breaking,
we found that it is possible to construct models that satisfy all of these conditions by considering a UV complete theory.
We also found that the predicted upper bound on the gluino mass
is within the reach of the LHC.
Therefore,  models with a gravitino mass in the eV range based on one-loop
R-symmetry breaking are expected to be supported or disproved in the near future.

Several final remarks are in order.
A gravitino with a mass in the eV range cannot be the dominant
component of the dark matter and these models require other dark matter candidates.
In models with gauge mediation, we have several places to look for a dark matter candidate. Dark matter candidates could possible be found in
the messenger or SUSY breaking sectors.
In fact, these sectors could provide stable particles with masses in the hundreds of TeV that could be interesting candidates for dark matter, since the dark matter density
can be consistent with observations if the annihilation cross sections
of the dark matter candidates saturate the unitarity bound.

For example, the messengers can be good candidates for dark matter
if they are confined into a composite state by strong dynamics and are neutral under the
SSM gauge groups.
In cascade gauge mediation, those composite states are expected
to have mass around 100 TeV, since the messenger scale is in that range.
The annihilation cross section of the composite neutral state
is expected to saturate the unitarity limit due to the strong dynamics responsible for their compositeness.
As discussed in \cite{Hamaguchi:2008rv,Hamaguchi:2009db},
models with composite messenger dark matter can be realized
for $N_{\rm mess}=5$ where the $SU(N_{\rm mess})$ symmetry
is identified with the strongly coupled gauge theory which is responsible
for creating the composite states out of the messengers.
The model with $N_{\rm mess}=5$ also has the feature of a relatively heavy gluino mass (see \eqrefc{eq:gluinoOPT}).

Another possibility is that the primary SUSY breaking sector
may provide dark matter candidates with masses in the TeV range in the form of pseudo-Nambu-Goldstone bosons.
As discussed in \cite{Ibe:2009dx,Ibe:2009en},
a consistent dark matter density is achieved through a Breit-Wigner enhancement\,\cite{Ibe:2008ye}
of the dark matter annihilation cross section.

In our argument, we have assumed that the secondary field, $S$, couples 
to $SU(3)$-triplet and $SU(2)$-doublet messengers universally.
If we introduce two secondary fields, $S_{d,\ell}$, which couple
to the triplet and doublet messengers separately,
the usual mass spectrum for SUSY particles
in minimal gauge mediation could be changed.
In such a case, however, one may worry about introduction of new CP-phases.
As discussed in \cite{Evans:2010ru}, however, there is no CP-problem even in this extension.

Finally, we note that the cascade gauge mediation model based on
the R-symmetry breaking models in this paper can also be used to construct
models with much heavier gravitinos.
One important application of such a possibility is to models in which a gravitino of mass 1-10 keV can be the dominant dark matter component.%
\footnote{A gravitino mass in the one to ten keV range is still difficult to realize in the conventional cascade gauge mediation models, where more than one loop is used to generate $m_S^2$ as in \cite{Dine:1993yw,Dine:1994vc,Dine:1995ag}.
}
Gravitino dark matter with mass in this range draws attention
as a possible solution to the seeming discrepancies between the observation
and the simulated results of galaxy formation based on the cold dark matter model\,\cite{WDM}.
The thermally produced gravitino density, however, exceeds the current observed
dark matter density.
Thus, we need to dilute the dark matter density by entropy production, for example, 
from the long lived particles in the primary SUSY breaking sector\,\cite{Ibe:2010ym}.

Another interesting feature of the cascade gauge mediation for a heavier
gravitino mass is the direct R-axion search at the LHC.
As we discussed in section\,\ref{sec:Raxion}, the mass of the R-axion increases with the
gravitino mass. The R-axion decay constant, which governs
the R-axion interactions, however, remains around $10^5/h$\,GeV even for a heavier
gravitino.
As discussed in \cite{Goh:2008xz}, the R-axion with a mass of hundreds of MeV can dominantly
decay into a pair of muons.
At the LHC, an R-axion with a decay constant $f_R = 10^{4-5}$\,GeV
can be copiously produced, and hence,
it is possible to detect the R-axion at the LHC by searching for the displaced vertex of the R-axion decay
from which a muon pair with low invariant mass is produced.

\section*{Acknowledgements}
This work was supported by the World Premier International Research Center Initiative
(WPI Initiative), MEXT, Japan.
\appendix
\section{Necessary Condition for R-symmetry Breaking at One Loop}\label{sec:NCR}
In this appendix, we show the need for fields with R-charge other than 0 or 2 to achieve
spontaneous R-breaking in O'Raifeartaigh models at one loop and at leading order in
SUSY breaking.
In our discussion we assume that R-symmetry is broken by the vev of an R-charged superfield $S$,
which is not necessarily identified with the SUSY breaking field $Z$ that obtains $\vev{F_Z} \neq 0$.
We also assume that the R-breaking field has no mass term in the superpotential, which
is the natural assumption since we are seeking models with R-symmetry breaking at one loop.

At leading order in SUSY breaking,
the soft mass of $S$ is given by,
\begin{eqnarray}
\label{eq:smassL}
 m_S^2 = \frac{c}{m^2} |F_Z|^2\ ,
\end{eqnarray}
where $c$ is the one-loop coefficient and $m$ is the mass scale of the fields that are integrated out.
For $m_S^2<0$, the scalar component of $S$ is repelled from its origin, and hence,
R-symmetry is spontaneously broken.

\begin{figure}[tbp]
\begin{center}
\includegraphics[width=0.5\textwidth]{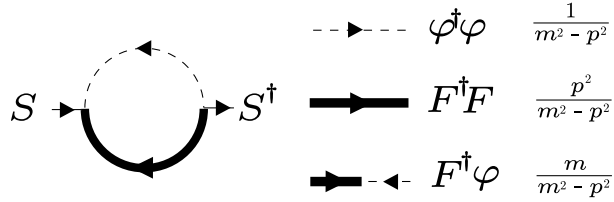}
\end{center}
\caption{
The diagram of the one-loop self energy of $S$ to which $|F_Z|^2$ is inserted.
Here, the dashed line denotes the scalar propagator, while the thick line denotes
that of the $F$-component.
}
\label{fig:fig2}
\end{figure}

To be specific let us assume that $S$ and $Z$
couple to some other superfields $X_i$ in the O'Raifeartaigh model%
\footnote{The $X_i$ may include $S$ or $Z$ as long as the stability of the SUSY breaking
vacuum is intact.}
via the superpotential terms,
\begin{eqnarray}
  y_{ij} S X_i X_j, \quad g_{ij} Z X_i X_j\ .
\end{eqnarray}
Here we assume that the $X_i$ have diagonalized mass terms at $S =Z = 0$.
The leading soft mass squared in \eqrefc{eq:smassL} is generated
by inserting $F_Z$ and $F_Z^*$ into the one-loop self-energy diagram of $S$ in which
only scalars are circulating (see Figure\,{\ref{fig:fig2}}).

\begin{figure}[tbp]
\begin{center}
\includegraphics[width=0.5\textwidth]{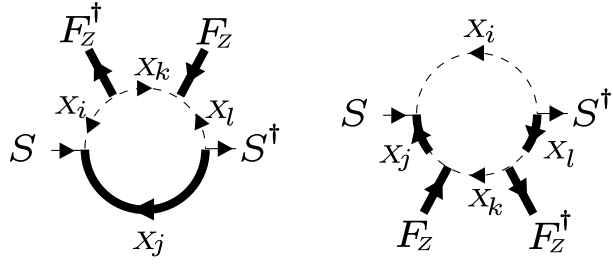}
\end{center}
\caption{
The diagrams with $|F_Z|^2$ insertion.
$|F_Z|^2$ is inserted on the scalar line on the left diagram, while it is inserted on the
$F$-term line.
}
\label{fig:fig3}
\end{figure}

From the figure, we see that there are two ways to insert $|F_Z|^2$ into the self-energy diagram.
One way is to insert $F_Z$ and $F_Z^*$ into the upper half of the loop, (i.e. the scalar line,
see Figure\,\ref{fig:fig3}). The contribution to $m_S^2$ is
\begin{eqnarray}
 m_{S,\phi}^2 = \frac{1}{16\pi^2}\sum_{i,j,k,l} \int\frac{d^4\ell_E}{(2\pi)^4}
y_{ji}
\frac{1}{m_{X_i}^2 + \ell_E^2}
g_{ik}^*
  \frac{1}{m_{X_k}^2 + \ell_E^2}
  g_{kl}
   \frac{1}{m_{X_l}^2 + \ell_E^2}
  y^{*}_{l j}
    \frac{1}{m_{X_j}^2 + \ell_E^2}\ell_E^2\ ,
\end{eqnarray}
where we have rotated the loop momentum to the Euclidian momentum.
Now, let us define,
\begin{eqnarray}
 X_{jk} &=& \sum_{i}\frac{1}{\sqrt{m_{X_j}^2+\ell_E^2}}y_{ji}
 \frac{1}{m_{X_i}^2 + \ell_E^2} g_{ik}^*
  \frac{1}{\sqrt{m_{X_k}^2 + \ell_E^2}}\ , \cr
 X^\prime_{km} &=& \sum_{l}\frac{1}{\sqrt{m_{X_k}^2+\ell_E^2}}g_{kl}
 \frac{1}{m_{X_l}^2 + \ell_E^2} y_{lm}^*
  \frac{1}{\sqrt{m_{X_m}^2 + \ell_E^2}} = X_{mk}^*\  ,
\end{eqnarray}
where we have used $y_{ij} = y_{ji}$ and $g_{ij}=g_{ji}$ to show that $X^\prime_{km}=X^*_{mk}$.
By using these matrices, the contribution to the mass is rewritten as
\begin{eqnarray}
 m_{S,\phi}^2 = \frac{1}{16\pi^2}\sum_{j,k} \int\frac{d^4\ell_E}{(2\pi)^4}
 X_{jk}X_{kj}^\prime\ell_E^2
 =\frac{1}{16\pi^2}\sum_{j,k} \int\frac{d^4\ell_E}{(2\pi)^4}
|X_{jk}|^2\ell_E^2
\ > 0\ .
\end{eqnarray}
Therefore, the contribution from the $|F_Z|^2$ insertion into the scalar line is always positive,
and hence, cannot cause R-symmetry breaking.

The other way to insert $|F_Z|^2$ is into the lower half of the loop (i.e. the $F$-line).
The resultant contribution to $m_S^2$ is given by,
\begin{eqnarray}
 m_{S,F}^2 =- \frac{1}{16\pi^2}\sum_{i,j,k,l} \int\frac{d^4\ell_E}{(2\pi)^4}
y_{ji}
\frac{1}{m_{X_i}^2 + \ell_E^2}
y_{il}^*
  \frac{1}{m_{X_l}^2 + \ell_E^2}
  g_{lk}^*
   \frac{1}{m_{X_k}^2 + \ell_E^2}
  g_{k j}
    \frac{1}{m_{X_j}^2 + \ell_E^2}m_j^* m_l \ .
\end{eqnarray}
This can be rewritten by using
\begin{eqnarray}
 Y_{ik} &=& \sum_{l}\frac{1}{\sqrt{m_{X_i}^2+\ell_E^2}}y^*_{il}
 \frac{m_l}{m_{X_l}^2 + \ell_E^2} g_{lk}^*
  \frac{1}{\sqrt{m_{X_k}^2 + \ell_E^2}}\ , \cr
 Y^\prime_{km} &=& \sum_{j}\frac{1}{\sqrt{m_{X_k}^2+\ell_E^2}}g_{kj}
 \frac{m_j^*}{m_{X_j}^2 + \ell_E^2} y_{jm}
  \frac{1}{\sqrt{m_{X_m}^2 + \ell_E^2}} = Y_{mk}^*\  ,
\end{eqnarray}
to
\begin{eqnarray}
 m_{S,F}^2 =- \frac{1}{16\pi^2}\sum_{i,k} \int\frac{d^4\ell_E}{(2\pi)^4}
| Y_{ik}|^2 \  <  0 \ .
\end{eqnarray}
Thus, the contribution from the  insertion to the $F$-line gives a negative contribution to $m_S^2$,
which may lead to R-symmetry breaking.

From this argument,
we see that the necessary condition to have negative $m_S^2+m_{S,\phi}^2+m_{S,F}^2$
at the leading order is that  the model allows the insertion of $|F_Z|^2$ to the internal $F$-line.
By cutting the diagram in between $|F_Z|^2$ on the $F$-line in Figure \ref{fig:fig3},
we see that the insertion to the $F$-line is possible when the effective term
\begin{eqnarray}
{\cal L}_{\rm eff} \propto  F_Z S X_i X_k  + h.c. \ ,
\end{eqnarray}
is allowed by the symmetries of the model, which corresponds to the term in the superpotential,
\begin{eqnarray}
{W}_{\rm eff} \propto  Z S X_i X_k\ .
\end{eqnarray}
Since we assume that the R-charge of $Z$ is $2$ and $S$ has a non-vanishing R-charge,
the above effective superpotential is allowed only when $S$ or $X_i$ or $X_k$
have R-charges other than $0$ or $2$.
This proves that R-symmetry breaking at one loop (and at leading order in $|F_Z|^2$)
requires the superfields with R-charge other than $0$ or $2$.

\section{$D$-term Spontaneous R-Symmetry Breaking}
\label{sec:Dterm}

Here we show that our desired one-loop R-symmetry breaking may be
induced by $D$-term SUSY breaking.  We will begin with a simple FI model with two flavors with charge of unit magnitude, $\phi_\pm$, $\widetilde\phi_\pm$.  Consider the superpotential
\begin{equation}
\label{eq:modelD}
W=m(\phi_+\widetilde\phi_-+\wt\phi_+\phi_-)+\lambda S \phi_+\phi_-\ +\dots,
\end{equation}
where the ellipsis stands in for $S$-independent, higher dimension terms, which do not affect the result. With our normalization conventions, $D$ takes the form
\begin{equation}
D=-\xi-g\left(|\phi_+|^2-|\phi_-|^2+|\wt\phi_+|^2-|\wt\phi_-|^2\right)\ ,
\end{equation}
where $g$ is the gauge coupling constant and $\xi$ is the mass-dimension-two FI parameter.
For this model, the one-loop mass for $S$ is found to be
\begin{eqnarray}
\label{eq:massSQD}
m_S^2 &=&\frac{\lambda^2}{16\pi^2}\left(\frac{g\xi}{m}\right)^2h(x)\ ,\cr\cr
h(x)&=&\frac{(2+x)\ln(1+x)+(2-x)\ln(1-x)}{x^2}\ ,
\end{eqnarray}
where we have taken $x=g\xi/m^2<1$.

\begin{figure}[tbp]
\begin{center}
\includegraphics[width=0.6\textwidth]{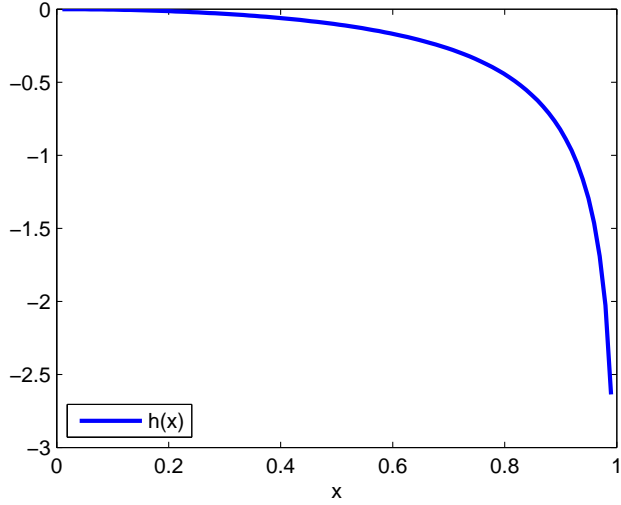}
\end{center}
\caption{
The function $h(x)$ ($x = ke/2m^2$)
given in \eqrefc{eq:massSQD}, which represents
the size of the tachyonic mass in the model based on the D-term SUSY breaking.
The value of $|h(x)|$ is highly enhanced in the region $x\to 1$,
which corresponds to an IR singularities in the one-loop diagram.
}
\label{fig:fig4}
\end{figure}

As we see from Figure\,\ref{fig:fig4}, $m_S^2$ is negative for all $x$, and so
R-symmetry breaking is achieved when $S$ has non-zero R-charge.
The disadvantage of this model is that the one-loop mass vanishes to leading order in $x$. This means that the scale of R-symmetry breaking is suppressed compared with the models
discussed in the main text unless $x$ is close to one.

The result is, in fact, singular in the limit $x\to1$.  This is an IR singularity that emerges when two of our fields become massless.  The diagram responsible is the one built from the trilinear scalar couplings,
\begin{equation}
\lambda m^*\phi_-\wt\phi_-^*S+h.c.
\end{equation}
Notice that, in the parameter region near the singularity, the estimate of the vacuum expectation value of $S$ in terms of $m_S$ in \eqrefc{eq:vevSapp} is no longer reliable. Instead we need to consider the full Coleman-Weinberg potential to determine the vacuum expectation
value of $S$, where the IR singularity is automatically cutoff by the VEV of $S$.
Thus, even in the parameter region where $|m_S|$ is highly enhanced by the IR singularity,
it is not expected that the vacuum expectation value of $S$ is highly enhanced from
\begin{equation}
\vev S \simeq \frac{\lambda^2}{16\sqrt{2}\pi^2\,h}\left(\frac{g\xi}{m}\right)^2\ .
\end{equation}

From this result, it follows that R-symmetry breaking of this sort occurs in a much broader class of models.  For example, we can consider couplings of the form in \eqref{eq:modelD}, but add an arbitrary number of fields with arbitrary (non-anomalous) charge assignments.  For our purposes, the model effectively decouples into a sum over sectors of like charge magnitude, $q$.  Each has the form,
\begin{equation}\label{DRB}
W_q=M_{ij}\phi_{+i}{\phi}_{-j} +\lambda_{ij}S\phi_{+i}\phi_{-j}\,.
\end{equation}
We further impose the constraint that there be an R-symmetry under which $S$ is charged.
\begin{equation}
M_{ij}\ne 0 \Rightarrow R(q_i)+R(\bar{q}_j)=2,\quad \lambda_{ij}\ne 0 \Rightarrow R(q_i)+R(\bar{q}_j)=\alpha,\quad R(S)=2-\alpha\ne0.
\end{equation}
$\alpha$ is undetermined here%
\footnote{Upon including the cubic term, $\Delta W\sim S^3$, it is fixed to $\alpha=4/3$.}.
As before, we also require $\vev{\phi_{\pm i}}=0$.  The matrix, $M$, may be diagonalized by a biunitary transformation leaving $D$ invariant.  It is then easy to see that the theory has a further effective decoupling.  If we consider the contribution to the mass of $S$ from a particular coupling $\lambda_{\alpha\beta}$ (Greek indices will not be summed) the relevant terms involve just two flavors.
\begin{eqnarray}
&&W\supset m_\alpha\phi_{+\alpha}\phi_{-\alpha}+m_\beta\phi_{+\beta}\phi_{-\beta}+\lambda_{\alpha\beta}S\phi_{+\alpha}\phi_{-\beta}\qquad\mbox{(no sum)},\cr
&&D\supset-\xi-qg(|\phi_{+\alpha}|^2-|\phi_{-\alpha}|^2+|\phi_{+\beta}|^2-|\phi_{-\beta}|^2)\,.
\end{eqnarray}
This means that we may directly apply the previous result by rescaling the coupling and summing over charges%
\footnote{We previously considered the case with $m_\alpha=m_\beta$. However, the one-loop mass with $m_\alpha\ne m_\beta$ generically gives $m_S^2<0$ as well.}.  
The upshot is that, as before, $S$ is tachyonic throughout the parameter space.

We may likewise extend this result to models with non-abelian D-term breaking \cite{nonfi}.  We still take $S$ to have no tree-level mass term, so it is not permitted to couple to the fields responsible for D-term breaking (and Higgsing).  For our purposes, these fields only play a role by contributing to the masses of the fields that do couple to $S$:
\begin{equation}
V\supset\vev{D^a}\sum_r\phi_r^\dagger T_r^a\phi_r\,.
\end{equation}
For example, in SUSY QCD, we may consider the analogous set of couplings to those in \eqref{DRB}, but now with $\phi_+$ a fundamental and $\phi_-$ an anti-fundamental under an $SU(N)$ gauge group.  After a set of field transformations, one finds that the squared mass of $S$ is given as a sum over colors of negative contributions, much like before.

\end{document}